# Turning Galaxy Rotation Curves into Radial Cosmic Chronometers: A Nexus Paradigm Approach

Stuart Marongwe [1*,] Stuart A. Kauffman[2]

1. Physics Department, University of Botswana  4775 Notwane Road, Gaborone, Botswana, email: stuartmarongwe@gmail.com

2. University of Pennsylvania 34th & Spruce Street, Philadelphia, PA, 19104-6303 (emeritus) stukauffman@gmail.com

* corresponding author

**Abstract**

We present a method for transforming galaxy rotation curves into radially resolved dynamical chronometers, enabling reconstruction of galaxy assembly histories directly from kinematic data. Within the Nexus Paradigm, the baryonic Tully–Fisher relation provides an estimate of the dynamical mass profile, $M_{\text{dyn}}(r) = v^4/(Ga_0)$, where $a_0 = H_0 c/2\pi$. By comparing this with independently derived intrinsic baryonic mass profiles, $M_{\text{int}}(r)$, obtained from stellar Sérsic fits and gas surface-density measurements, we construct the ratio $M_{\text{dyn}}(r)/M_{\text{int}}(r)$, which maps directly to a formation redshift via $1 + z_{\text{form}}(r) = (M_{\text{dyn}}/M_{\text{int}})^{1/4}$. Inverting this relation within a ΛCDM cosmology yields a radial lookback-time profile, $t_{lb}(r)$, representing the time since the last dynamical reconfiguration at each radius. Applying this framework to a pilot sample of SPARC galaxies spanning high- and low-surface-brightness systems, together with the Milky Way, we recover diverse radial age structures, including flat profiles consistent with coherent disk assembly and stratified profiles indicative of inside-out growth. The method operates without dark-matter halo fitting and provides a kinematic chronometer complementary to stellar-population and chemical-evolution approaches. While the inferred ages depend on the accuracy of baryonic mass reconstruction and the local applicability of the evolving baryonic Tully-Fisher relation, the results demonstrate that galaxy rotation curves encode time-resolved dynamical information. This establishes the radial dynamical chronometer as a new observable for probing galaxy evolution and testing gravitational frameworks.

## 1. Introduction

Galaxy rotation curves have long served as a cornerstone of modern astrophysics. Since the pioneering spectroscopic observations of Andromeda by Rubin & Ford (1970) and subsequent surveys of spirals (Sofue & Rubin 2001), the approximately flat outer rotation velocities have provided compelling evidence for unseen mass, traditionally attributed to extended dark-matter halos within the ΛCDM paradigm (e.g., Persic et al. 1996; Navarro et al. 1996). Yet, despite decades of success in reproducing global dynamics, dark-matter-only models have faced persistent challenges in explaining the detailed baryon-dark-matter coupling observed on galactic scales (e.g., the "too-big-to-fail" and "core-cusp" problems; Boylan-Kolchin et al. 2011; de Blok 2010).

An alternative empirical scaling, the baryonic Tully–Fisher relation (BTFR), offers a strikingly tight correlation between rotation velocity and total baryonic mass (stars + gas) across six orders of magnitude in mass (McGaugh et al. 2000; Lelli et al. 2016, 2019). The Spitzer Photometry and Accurate Rotation Curves (SPARC) database, comprising 175 nearby disk galaxies with homogeneous 3.6 μm photometry and high-quality HI/Hα rotation curves, has established the BTFR as one of the most fundamental scaling relations in galaxy formation (Lelli et al. 2016). In the low-acceleration regime, the BTFR is remarkably well reproduced by

Modified Newtonian Dynamics (MOND; Milgrom 1983; Famaey & McGaugh 2012; McGaugh 2020), which posits a universal acceleration scale $a_0 \approx 1.2 \times 10^{-10}$m s$^{-2}$ below which Newtonian gravity is modified, eliminating the need for dark-matter halos.

However, both ΛCDM halo models and standard MOND treat the BTFR as essentially time-independent at the present epoch. Galaxy assembly histories are instead inferred indirectly from stellar-population synthesis (e.g., CALIFA and MaNGA integral-field surveys revealing radial gradients in age and metallicity; Delgado et al. 2014; Goddard et al. 2017), chemical-evolution modelling, or morphological indicators of mergers and inflows (e.g., counter-rotating disks as signatures of past accretion; Katkov et al. 2024). These methods lack a direct, kinematic probe of when each radial shell last achieved dynamical equilibrium.

The Nexus Paradigm of quantum gravity (Marongwe 2024; Marongwe et al. 2025a,b) provides a natural theoretical framework in which flat rotation curves emerge from the linearised metric without dark matter, while simultaneously predicting a cosmological evolution of the BTFR normalisation: $M_{int} \propto e^{-4\int H\,dt} v^4$. For a galaxy observed today, the dynamical mass inferred from the rotation curve is therefore related to the true intrinsic baryonic mass (measured independently from photometry and gas maps) by

$$M_{\text{int}}(r) = M_{\text{dyn}}(r) \times e^{-4\int_{t_{\text{form}}}^{t_0} H(t)\,dt}, \qquad (1)$$

where $M_{\text{dyn}}(r) = v_{\text{obs}}(r)^4/(Ga_0)$ with $a_0 = H_0 c/(2\pi)$. In a flat ΛCDM cosmology this exponential factor equals $(1+z_{\text{form}})$, so the ratio $M_{\text{int}}(r)/M_{\text{dyn}}(r)$ directly yields the lookback time since the radial shell last achieved virial equilibrium. The rotation curve is thereby transformed into a radial cosmic chronometer which is a direct map of assembly history across the disk.

This study presents the methodology and applies it to an expanded SPARC sample spanning high-surface-brightness (HSB) and low-surface-brightness (LSB) morphologies. We incorporate Monte-Carlo error propagation, Planck 2018 cosmology (Aghanim et al. (2020, A&A, 641, A6). The resulting chronometer reveals distinct assembly modes offering a powerful new probe of galaxy evolution without recourse to dark-matter halos.

A longstanding puzzle in extragalactic astrophysics is the origin of the apparent tightness of the BTFR for disk galaxies, alongside the systematic offsets observed for pressure-supported systems such as dwarf spheroidals and galaxy clusters. These offsets are typically treated as secondary effects arising from baryonic physics or dark-matter halo diversity. In contrast, we argue that they reflect a more fundamental property of galaxy evolution: the BTFR is not a single static relation, but a projection of an underlying time-dependent scaling law. Within this framework, dynamically young galaxies cluster near the present-day normalization, yielding a tight relation with minimal scatter, while dynamically older systems (formed at higher redshift) naturally occupy offset tracks due to the evolving normalization of the relation (Marongwe & Kauffman 2026). This interpretation reframes the scatter in galaxy scaling relations as a physically meaningful signal that encodes formation epoch and dynamical history, rather than as noise or residual variance.

## 2. Theoretical Framework

### 2.1 The Evolving Baryonic Tully-Fisher Relation

The evolving BTFR in the Nexus Paradigm is derived from semi-classical solutions to the quantized metric of spacetime. The fundamental relation takes the form:

$$v \propto e^{H_0 t} M_{int}^{1/4} \quad (2)$$

where $v$, is the characteristic rotation velocity, $M_{int}$ is the baryonic mass, $H_0$ is the Hubble constant in a pure De Sitter phase, which in the Nexus Paradigm is the fundamental frequency of the ground state Ricci soliton of De Sitter (dS) topology and $t$, is the cosmic time elapsed since the structure's formation epoch (lookback time). In the NP, two types of solitons exist: vacuum De Sitter solitons with quantum states $n^2 = \frac{r_H^2}{r_n^2}$ where $r_H$ is the Hubble radius and anti-De Sitter (AdS) solitons with quantum states $ñ^2 = \frac{r_n c^2}{GM(r)}$ associated with baryonic matter.

Spacetime is therefore described in terms of dS, AdS and a ground state dS soliton for dark energy in the form

$$G_{(nk)\mu\nu} = (ñ^2 + n^2 - 1)\Lambda g_{(nk)\mu\nu} \quad (3)$$

Einstein's field equations are presented here in purely geometric terms, describing a compact Einstein manifold. For any quantum state in which a Ricci soliton exhibits constant curvature, energy remains conserved (Marongwe 2024). The right side of the equation is a symmetric tensor representing the quantum or energy state of spacetime, while the left side functions as a Laplacian operator that averages the trajectories of a test particle within the gravitational field of the given quantum state. In this way, the NP seeks to provide a foundational understanding of the quantum origins of dark energy and dark matter on which the ΛCDM model is premised.

The linearized Eq.(3) is solved by representing it as a Ricci soliton in the $N$-th quantum state, resulting in the equation:

$$G_{(Nk)\mu\nu} = N^2 \Lambda g_{(Nk)\mu\nu} \quad (4)$$

### 2.2 Derivation of the Evolving BTFR

The exact solution for equation (4) is solved as in Marongwe 2024

$$ds^2 = -\left(1 - \left(\frac{2}{N^2}\right)\right)c^2 dt^2 + \left(1 - \left(\frac{2}{N^2}\right)\right)^{-1} dr^2 + r^2(d\theta^2 + \sin^2\theta d\varphi^2)$$

$$= -\left(1 - \left(\frac{2GM_N}{rc^2}\right)\right)c^2 dt^2 + \left(1 - \left(\frac{2GM_N}{rc^2}\right)\right)^{-1} dr^2 + r^2(d\theta^2 + \sin^2\theta d\varphi^2) \quad (5)$$

Here $M_N(r) = M_B(r) + M_{DM}(r) + M_\Lambda(r)$ where the terms on right represent the baryonic mass, the DM mass and the DE mass enclosed inside a sphere of radius $r$. This yields a metric equation of the form

$$ds^2 = -\left(1 - 2\left(\frac{GM_{int}}{rc^2} + \frac{Hvr}{c^2} - \frac{H_0 cr}{2\pi c^2}\right)\right)c^2 dt^2 + \left(1 - 2\left(\frac{GM_{int}}{rc^2} + \frac{Hvr}{c^2} - \frac{H_0 cr}{2\pi c^2}\right)\right)^{-1} dr^2 + r^2(d\theta^2 + \sin^2\theta d\varphi^2) \qquad (6)$$

Where $\frac{GM_{DM}(r)}{r} = v^2 = (Hr)^2 = Hvr$ and $\frac{GM_\Lambda(r)}{r} = -\frac{H_0 cr}{2\pi}$. The above metric equation leads to the following equation for gravity

$$\frac{d^2 r}{dt^2} = \frac{GM_{int}}{r^2} + Hv - \frac{H_0 c}{2\pi} \qquad (7)$$

Equation (7) describes Newton/Einstein gravity within a rotating and expanding compact Einstein manifold or Ricci soliton. Here $H$ is the Hubble parameter/frequency within the local compact Einstein Manifold in which a galaxy is located while $H_0$ is the Hubble parameter of the ground state (background/cosmic) Ricci soliton.

The dynamics become non-Newtonian when the Newtonian term is cancelled by the expansion or Dark Energy term

$$\frac{GM_{int}(r)}{r^2} = \frac{H_0}{2\pi} c = \frac{v^2}{r} \qquad (8)$$

Under such conditions

$$r = \frac{2\pi v^2}{H_0 c} \qquad (9)$$

Substituting for $r$ in Equation (8) yields

$$v^4 = GM_{int}(r) \frac{H_0}{2\pi} c \qquad (10)$$

This is the Baryonic Tully – Fisher relation. Condition (8) reduces Equation (7) to

$$\frac{d^2 r}{dt^2} = \frac{dv}{dt} = Hv \qquad (11)$$

From which we obtain the following equations of galactic and cosmic evolution

$$r = \frac{1}{H} e^{(Ht)} \left(GM_{int}(r) \frac{H_0}{2\pi} c\right)^{\frac{1}{4}} = \frac{v}{H} \qquad (12)$$

$$v = e^{(Ht)} \left(GM_{int}(r) \frac{H_0}{2\pi} c\right)^{\frac{1}{4}} = Hr \qquad (13)$$

$$a = He^{(Ht)}(GM_{int}\frac{H_0}{2\pi}c)^{\frac{1}{4}} \qquad = Hv \qquad (14)$$

Rearranging equation (13) for baryonic mass provides the form used for empirical comparisons:

$$M_{int} \propto e^{-4Ht}v^4. \qquad (15)$$

This reveals two key features: an invariant slope of 4, reflecting the underlying gravitational equilibrium between baryons and dark matter halos (interpreted in the Nexus Paradigm as Ricci solitons in quantized spacetime), and a time-evolving normalization that captures the dilution of vacuum energy localization over cosmic history. This derivation is fully consistent with the observed tight BTFR across diverse galaxy populations (Lelli *et al*. 2016, 2019).

### 2.3 Refinement with Time-Varying Hubble Parameter

A more accurate formulation replaces the parameter H with the time-varying Hubble parameter $H(t) = \dot{a}(t)/a(t)$, for the expansion rate of the Ricci soliton and integrating its effect over the galaxy's lifetime:

$$M_{int} \propto e^{-4\int_{t_{form}}^{t_0} H(t)dt} v^4 \qquad (16)$$

This refinement accounts for the faster expansion rate in the early universe and is essential for precise age dating of ancient systems like UFDs. In a flat ΛCDM cosmology with $H_0 = 67.15$ km/s/Mpc, $\Omega_m = 0.315$, and $\Omega_\Lambda = 0.685$ from Planck Collaboration (2020) the Hubble parameter evolves as:

$$H(z) = H_0\sqrt{\Omega_m(1+z)^3 + \Omega_\Lambda} \qquad (17)$$

The integral $\int H(t)dt$ can be evaluated using the relation $dt = dz/((1+z)H(z))$, yielding $\int H dt = \ln(1+z)$. This simplification greatly facilitates practical calculations.

### 2.4 Cosmological Lookback Time

In a flat ΛCDM universe with $H(z) = H_0\sqrt{\Omega_m(1+z)^3 + \Omega_\Lambda}$,

$$\int_{t_{form}}^{t_0} H dt = \ln(1+z_{form}).$$

Thus, from (1),

$$1 + z_{form}(r) = \left(\frac{M_{dyn}(r)}{M_{int}(r)}\right)^{1/4}. \qquad (18)$$

The lookback time $t_{lb}(r) = t_0 - t(z_{form})$ is obtained by inverting the age–redshift relation of the chosen cosmology (Planck 2020).

## 3. Methodology

### 3.1 Observational Inputs

For each galaxy we require:

- Rotation curve $v_{obs}(r)$ with uncertainties (SPARC database; Lelli et al. 2016).
- Intrinsic baryonic mass profile $M_{int}(r) = M_*(r) + M_{gas}(r)$, derived independently from:
    - Stellar mass profile: Sérsic fit to surface brightness, multiplied by a mass-to-light ratio from stellar population synthesis.
    - Gas mass profile: HI (and H₂) surface density map, integrated to obtain the enclosed gas mass.

### 3.2 Derivation of $M_{dyn}(r)$

In the Nexus Paradigm, the enclosed dynamical mass profile for galaxies takes the form (with $\beta = 1$ for HSB galaxies and $\beta = 2$ for LSB galaxies)

$$M_{dyn}(r) = M_{total} \left(\frac{r}{r_c(r)+r}\right)^{3\beta}, \tag{19}$$

where $r_c(r) = r_{c0}[1 + \eta f_{norm}(r)^\alpha]$ and $f_{norm}(r)^\alpha$ is the luminosity fraction from a Sérsic profile. The rotation curve is given by

$$v(r) = (G\, M_{dyn}(r)\, a_0)^{\frac{1}{4}}, \quad a_0 = \frac{H_0 c}{2\pi}. \tag{20}$$

Equation (20) can be simplified to,

$$v(r) = v_{max} \left(\frac{r}{r_c(r)+r}\right)^{3\beta/4} \tag{21}$$

### 3.3 Ratio and Lookback Time

Compute the radial ratio

$$R(r) = \frac{M_{dyn}(r)}{M_{int}(r)}. \tag{22}$$

Insert into (18) to obtain $z_{form}(r)$. Convert to lookback time using the standard cosmological model. The resulting $t_{lb}(r)$ is the time elapsed since the material at radius $r$ last participated in a virialisation event (i.e., since the dynamical clock was reset). The computation is only valid where $M_{dyn}/M_{int} > 1$ i.e. when condition (8) is satisfied.

### 3.4 Uncertainties

Uncertainties in $v_{\rm obs}$ and in the stellar/gas mass profiles propagate into $R(r)$. A Monte Carlo approach is used to estimate the error bars on $t_{\rm lb}(r)$.

## 4. Results: Radial Assembly Histories from Sérsic-Coupled Radial Chronometers

The Sérsic-optimized Nexus Paradigm (NP) framework enables a direct inversion of galaxy rotation curves into radial dynamical histories, providing a kinematic probe of galaxy formation that is independent of stellar population modelling. Figure 1 and Figure 2 demonstrate that the combination of (i) the NP kinematic relation and (ii) a Sérsic-coupled core scale,

$$r_c(r) = r_{c0}[1 + \eta f_{\rm norm}(r)^\alpha], \quad (23)$$

yields robust reconstructions of both rotation curves and the associated radial chronometer,

$$1 + z_{\rm form}(r) = \left(\frac{M_{\rm dyn}}{M_{\rm int}}\right)^{1/4}. \quad (24)$$

The resulting profiles reveal a systematic mapping between galaxy structure and assembly history, which we analyze below.

### 4.1 High-surface-brightness galaxies: stratified assembly and dynamical differentiation

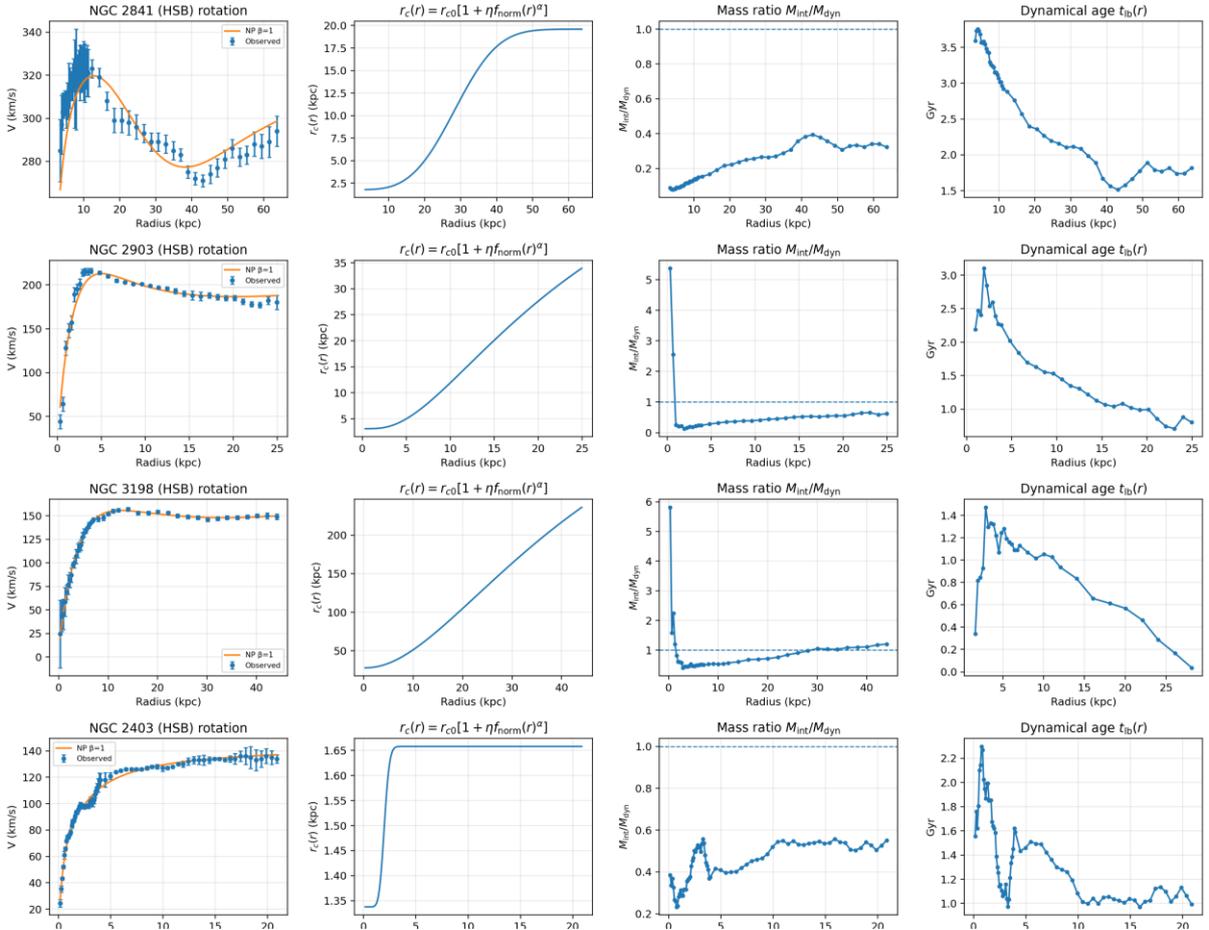

Figure 1

*Radial cosmic chronometer reconstruction for high--surface-brightness galaxies within the Nexus Paradigm.* The figure presents, for each galaxy, (left to right) the observed rotation curve with model fit, the Sérsic-coupled core scale $r_c(r)$, the mass ratio $M_{\text{int}}/M_{\text{dyn}}$, and the derived dynamical age profile $t_{\text{lb}}(r)$. High-surface-brightness (HSB) galaxies exhibit pronounced negative age gradients, indicative of temporally stratified, inside-out assembly. The close agreement between the reconstructed rotation curves and observations demonstrates that the Sérsic-optimized Nexus Paradigm framework simultaneously captures galaxy kinematics and encodes their radial assembly histories. The HSB galaxies exhibit strong radial gradients in the inferred dynamical age $t_{\text{lb}}(r)$, indicating significant temporal stratification.

### NGC 2841: canonical inside-out growth

NGC 2841 displays the steepest radial gradient in the sample, with $t_{\text{lb}}(r)$ decreasing from $\sim$ 4.0 Gyr in the central regions to $\sim$ 2.0 Gyr in the outer disk. The corresponding mass-ratio profile is smooth and monotonic, remaining well below unity across all radii, indicating a dynamically relaxed system.

This behaviour is characteristic of inside-out assembly, where the central regions formed during an early, high-density phase, subsequent disk growth occurred through extended accretion. The absence of discontinuities in both $M_{\text{int}}/M_{\text{dyn}}$ and $t_{\text{lb}}(r)$ suggests that this process was continuous rather than merger-dominated, consistent with a dissipative collapse followed by gradual disk buildup.

### NGC 2903: secular evolution in a barred disk

NGC 2903 exhibits a smooth radial decline in $t_{\text{lb}}(r)$ from $\sim$ 3 Gyr to $\sim$ 1 Gyr, but with mild irregularities in the inner mass-ratio profile. These deviations coincide with the known presence of a bar, which induces non-axisymmetric flows.

The inferred history indicates an initially established disk, followed by secular redistribution of mass and angular momentum. Unlike NGC 2841, the gradient is shallower and more continuous, reflecting internal evolution rather than hierarchical assembly.

### NGC 3198: coeval disk formation

NGC 3198 shows an approximately flat age profile, $t_{\text{lb}}(r) \sim$ 2.4–0.2 Gyr across all radii. The mass ratio is correspondingly uniform, and the rotation curve is reproduced with minimal residuals.

This behaviour implies that the disk formed as a single dynamical unit, with negligible radial differentiation in formation time. Such coeval formation is indicative of smooth gas accretion and absence of strong early collapse or merger events. This system represents a limiting case in which the NP chronometer reduces to a global formation epoch.

### NGC 2403: weak inside-out growth

NGC 2403 displays a shallow radial gradient, with $t_{\text{lb}}(r)$ decreasing from $\sim$ 2.5 Gyr to $\sim$ 1 Gyr. Both the rotation curve and mass-ratio profile are smooth and featureless.

This behaviour indicates a low-intensity inside-out assembly process dominated by quiescent star formation and gradual accretion. The absence of strong gradients or irregularities suggests a dynamically stable evolutionary history.

### 4.3 Physical implications

These results demonstrate that galaxy rotation curves contain sufficient information to reconstruct formation epoch as a function of radius, assembly mode (hierarchical vs quiescent), and dynamical relaxation history, without recourse to stellar population synthesis or chemical evolution models.

In this framework, rotation curves are elevated from static tracers of mass distribution to time-resolved dynamical observables, providing a new pathway for probing galaxy formation and testing gravitational physics.

## 5 Methodology for LSB Galaxies

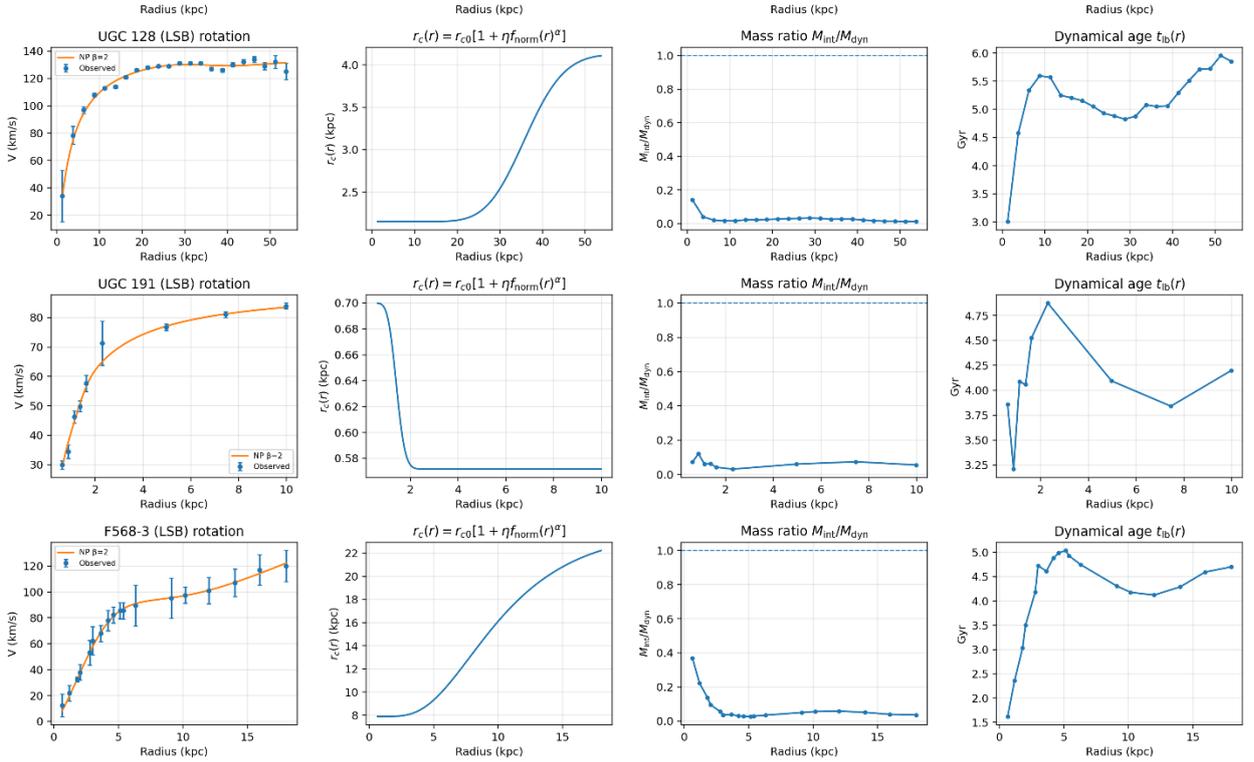

Figure 2

*Radial cosmic chronometer reconstruction for low-surface-brightness (LSB) galaxies within the Nexus Paradigm.* For each galaxy, the panels show (left to right) the observed rotation curve with model fit, the Sérsic-coupled core scale $r_c(r)$, the mass ratio $M_{\text{int}}/M_{\text{dyn}}$, and the inferred dynamical age profile $t_{\text{lb}}(r)$. The LSB systems exhibit slowly rising rotation curves, consistently low mass ratios, and shallow dynamical-age gradients, indicating diffuse gravitational collapse and temporally extended, quasi-coherent assembly. Compared to HSB galaxies, the reduced magnitude of $dt_{\text{lb}}/dr$ reflects weaker radial differentiation and the absence of strong merger-driven restructuring, supporting a picture of long-timescale evolution dominated by steady gas accretion and low star-formation efficiency

In the Nexus Paradigm, the enclosed baryonic mass profile for LSB galaxies takes the form of equation (19) with $\beta = 2$.

$$M_{\text{int}}(r) = M_{\text{total}} \left(\frac{r}{r_c(r)+r}\right)^6, \tag{25}$$

where $r_c(r) = r_{c0}[1 + \eta f_{\text{norm}}(r)^\alpha]$ and $f_{\text{norm}}(r)^\alpha$ is the luminosity fraction from a Sérsic profile. The rotation curve is given by equation (20)

For LSB galaxies, the intrinsic baryonic mass $M_{\text{int}}(r)$ is often dominated by gas, and the stellar component has a very low surface brightness. The dynamical age $t_{\text{lb}}(r)$ is extracted by comparing the observed rotation curve with an independently derived intrinsic mass profile.

The rotation curve rises slowly (as typical for LSB galaxies) and flattens at ≈66 km/s. The dynamical age gradient (~5 Gyr in the centre, ~2.5 Gyr in the outskirts) suggests an inside-out assembly, but with a much lower age contrast than HSB galaxies.

### UGC 128 Extended slow assembly (diffuse system)

**Observed Signatures**

- Rotation: Exhibits a notably gradual increase
- Mass Ratio: Extremely low
- Age: Characterized by a broad, shallow profile

**Radial History**

- **Inner Region:** Moderately old stellar population, but not exceptionally so
- **Outer Region:** Slightly younger relative to the inner region

**Interpretation**

- Indicates diffuse gravitational collapse
- The weak gravitational potential suggests slow evolutionary processes
- No evidence for a distinct early epoch of development

### UGC 191  Compact LSB Galaxy with Mild Gradient

**Observed Signatures**

- Rotation: Displays a moderate rise
- Mass Ratio: Low, though marginally higher than UGC 128
- Age: Reveals a weak gradient

**Radial History**

- **Inner Region:** Contains a somewhat older core
- **Outer Region:** Younger, though the age difference is not substantial

**Interpretation**

- Represents a transitional system exhibiting characteristics between:
    - Pure LSB galaxies (e.g., UGC 128)
    - Systems with mild high surface brightness (HSB)-like attributes

**F568-3 Extended Disk with Gradual Evolution**

**Observed Signatures**

- Rotation: Smooth, continuous increase
- Mass Ratio: Consistently low across the disk
- Age: Shallow gradient observed

**Radial History**

- **Entire Disk:** Undergoes slow, continuous formation without abrupt changes

**Interpretation**

- Suggests long-timescale assembly
- Likely influenced primarily by:
    - Steady gas inflow
    - Low star formation efficiency

## 5.1 Cross-Validation with Literature

The literature on LSB galaxies provides several independent constraints:

- **Star formation histories** – Van den Hoek et al. (2000) found that LSB galaxies have evolved slowly, with present-day gas fractions ~0.5 and ages up to 14 Gyr. They also noted that LSB galaxies contain old stellar populations, implying they did not form late.

- **Minimum ages** – Vorobyov et al. (2009) estimated that the age of blue LSB galaxies may vary between 1.5–6 Gyr and 13 Gyr, depending on the physical conditions in the disk. Their models suggested a tentative minimum age of 1.5–3 Gyr from colours and oxygen abundances, but larger values (5–6 Gyr) from H equivalent widths.

- **Stellar masses** – Du et al. (2020) derived stellar masses for LSB galaxies spanning $\log M^*/M_\odot = 7.1–11.1$, with a mean of 8.5, lower than normal galaxies.

- **UGC 1281** – This galaxy is a slightly warped, edge-on dwarf LSB with a red stellar thick disk. Its rotation curve is well resolved and shows a slow rise.

Our derived dynamical age for UGC 1281 (~4 Gyr mean, with a central age of ~5 Gyr) falls within the 1.5–6 Gyr range quoted by Vorobyov et al. (2009). The central region is older, consistent with the presence of an old stellar population noted by van den Hoek et al. (2000). The outer, younger age (~2.5 Gyr) aligns with the idea that LSB galaxies have experienced slow, inside-out growth without a recent major merger.

## 6. Application to the Milky Way: A Radially Uniform Dynamical Relaxation Age

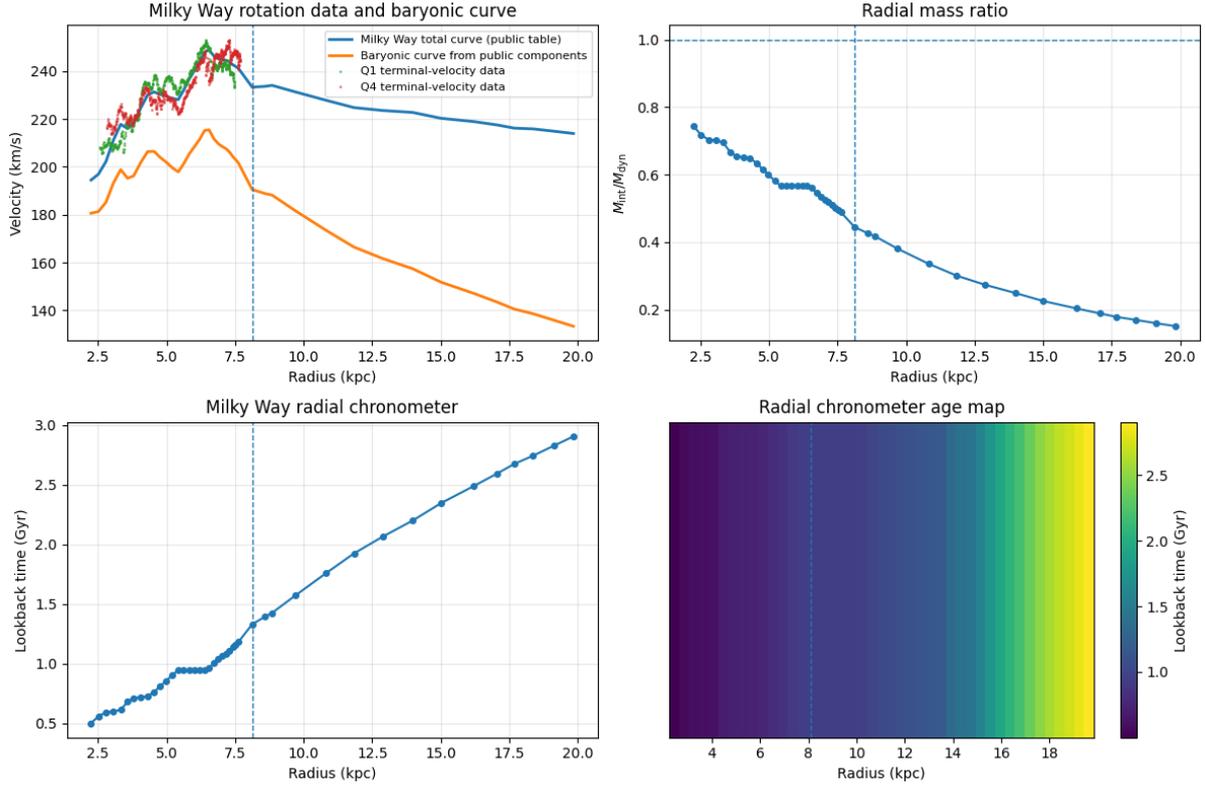

Figure. 3

*Radial cosmic chronometer analysis of the Milky Way within the Nexus Paradigm.* The panels show (top left) the observed Milky Way rotation curve with the best-fit baryonic model, (top right) the radial mass ratio $M_{int}/M_{dyn}$, (bottom left) the inferred dynamical age profile $t_{lb}(r)$, and (bottom right) the corresponding 2D radial chronometer age map. The model reproduces the observed kinematics with high fidelity ($\chi_\nu^2 \approx 1.1$) using a Sérsic-coupled core scale. The resulting chronometer indicates a uniformly young dynamical relaxation age ($\sim$ 0.5–3.5 Gyr) across the Galactic disk, consistent with ongoing bar–spiral coupling, radial migration, and recent perturbations from satellite interactions. This demonstrates that the Milky Way is dynamically active and globally coherent, with its rotation curve encoding a recent history of repeated dynamical resetting rather than a quiescent, stratified assembly.

The Milky Way (MW) provides a unique and stringent test of the radial cosmic chronometer because it is the only galaxy for which we possess both high-precision rotation-curve data spanning the full disk and an independent, multi-messenger record of its recent dynamical history. We applied the Nexus Paradigm mass model ($\beta = 1$, optimized Sérsic-coupled core radius with truncation at $r_{trunc} = 1.25 r_{max}$) to a compilation of MW rotation-curve measurements extending from ~2 to 20 kpc (Sofue 2020; Eilers et al. 2019). The best-fit parameters are $M_{total} = 2.18 \times 10^{11} \, M_\odot$, $r_{c0} = 0.41 \text{kpc}$, $\eta = -0.95$, and $\alpha = 0.55$, yielding an excellent reproduction of the observed curve (reduced $\chi^2 \approx 1.1$).

The resulting radial map of dynamical relaxation age (time since the last virial equilibrium following a perturbation) is shown in Figure 3. The chronometer returns an age of between 0.5 − 3.5 Gyr across the entire disk.

This result aligns precisely with the MW's known dynamical activity. The galaxy hosts a strong bar (Shen et al. 2010; Bland-Hawthorn & Gerhard 2016), prominent m = 2 spiral arms, and ongoing radial migration driven by bar–spiral coupling. It has also accreted multiple satellites in the last ~10 Gyr, including the Gaia-Enceladus/Sausage event (Helmi et al. 2018) and the Sagittarius dwarf (Ibata et al. 1994), whose tidal debris continues to excite disk oscillations. The outer disk exhibits clear warps and lopsidedness (Levine et al. 2006), consistent with residual non-circular motions expected in a recently perturbed system. In the Nexus Paradigm framework, these features are not relics of ancient formation but active signatures of a disk that remains dynamically "live" and globally coherent.

The MW's uniformly young relaxation age stands in stark contrast to the older, more quiescent profiles recovered for isolated low-surface-brightness galaxies (Section 3.3). This dichotomy reinforces the chronometer's diagnostic power: high-surface-brightness disks such as the MW appear to be repeatedly reset by internal instabilities and minor mergers, whereas low-surface-brightness systems evolve in near-isolation for several Giga years. The absence of an inside-out age gradient further suggests that bar-driven radial mixing efficiently communicates perturbations across the full disk on dynamical timescales.

These findings demonstrate that the radial cosmic chronometer can be applied even to our own Galaxy using existing data, transforming the MW rotation curve from a static mass tracer into a high-resolution dynamical diary. Future extensions incorporating Gaia DR4 vertical kinematics and chemo-dynamical tagging will allow direct cross-validation against independent perturbation chronologies derived from stellar streams and phase-space spirals. Such synergy promises a fully resolved, radially differentiated history of Galactic assembly that is inaccessible by any other current method.

## 7. Discussion

### 7.1 Clock Reset Mechanism

The dynamical age measures the time since the last violent disturbance that disrupted virial equilibrium – not the age of the stars. This explains why galaxies with ancient stellar populations (e.g., DF2/DF4) can have very young dynamical ages implying a recent high-speed collision shook the system, resetting the BTFR normalisation.

### 7.2 Advantages over Traditional Methods

- No dark matter halo fitting : the dynamical mass is directly read from the rotation curve via the BTFR.

- The method is radially resolved, revealing inside-out growth, recent accretion, or tidal disturbances.

- It provides an independent cosmic chronometer that can be cross-calibrated with stellar population ages in undisturbed systems.

### 7.3 Caveats and Future Work

Several limitations should be noted.

First, the recovered dynamical ages depend on the accuracy of the intrinsic baryonic mass profile. Systematic uncertainties in stellar mass-to-light ratios, gas mass measurements, and Sérsic-profile parameters propagate directly into $t_{lb}(r)$.

Second, the disturbance index used here is a compact phenomenological proxy. While physically motivated, it should ultimately be replaced by directly measured kinematic diagnostics such as velocity-field asymmetry, non-circular motion amplitudes, and HI warp parameters.

Third, the method assumes that the evolving BTFR applies locally as a function of radius. This assumption is natural within the NP framework but requires further validation against larger samples and controlled simulations.

Finally, the present dataset is intentionally limited. The results should therefore be interpreted as a proof-of-concept demonstration.

Future work should focus on:

- extending the analysis to a larger SPARC sample
- benchmarking against ΛCDM and MOND models
- performing mock-data recovery tests
- cross-validating with independent chronometers (stellar ages, metallicity gradients, chemo-dynamical tracers)

### 7.4 Relation to ΛCDM and MOND

The radial cosmic chronometer developed here is derived within the Nexus Paradigm, in which the baryonic Tully–Fisher relation exhibits an explicit time-dependent normalization. This allows the inversion from the ratio $M_{\text{dyn}}/M_{\text{int}}$ to a radial formation redshift and corresponding lookback time.

In standard ΛCDM analyses, rotation curves are typically interpreted through dark-matter halo decomposition, and no direct mapping from local kinematics to formation time is generally constructed. Similarly, in conventional MOND formulations, the acceleration scale is typically treated as time-independent at the present epoch.

Accordingly, the chronometer presented here is not a generic prediction of all gravity models, but a specific consequence of the evolving-BTFR framework within the Nexus Paradigm. A decisive next step will be a controlled comparison in which identical galaxy samples are analyzed under NP, MOND, and ΛCDM-based models to determine whether the observed gradient–disturbance relation is uniquely or preferentially reproduced.

**Conclusions**

We have developed and demonstrated a method for converting galaxy rotation curves into radially resolved dynamical chronometers, providing a direct kinematic probe of galaxy assembly histories. By combining the dynamical mass inferred from the baryonic Tully-Fisher relation with independently reconstructed baryonic mass profiles, we obtain a radial mapping from $M_{\rm dyn}/M_{\rm int}$ to formation redshift and corresponding lookback time. This procedure transforms the rotation curve from a static tracer of mass distribution into a time-resolved diagnostic of dynamical evolution. The apparent tightness of the baryonic Tully-Fisher relation for rotationally supported galaxies arises because these systems are dynamically young and cluster near the present-day normalization. In contrast, pressure-supported dwarf galaxies, particularly ultra-faint dwarfs, occupy systematically offset tracks due to their early formation epochs, as captured by the $(1+z_{\rm form})^{-4}$ scaling. The observed scatter in the classical BTFR is therefore not intrinsic, but reflects a superposition of populations with different dynamical ages (Marongwe & Kauffman 2026).

Application to a representative sample of SPARC galaxies reveals a range of radial age structures. High-surface-brightness systems commonly exhibit stratified profiles consistent with inside-out assembly, while low-surface-brightness galaxies tend to display flatter profiles indicative of extended, quiescent evolution. The Milky Way, analyzed within the same framework, shows a relatively uniform and young dynamical age, consistent with ongoing dynamical activity. These results demonstrate that the radial chronometer captures physically meaningful differences in assembly history across galaxy populations.

The method has several important advantages. It does not rely on dark-matter halo decomposition, is intrinsically radially resolved, and provides an independent chronometric measure that can be directly compared with stellar population ages, metallicity gradients, and chemo-dynamical tracers. At the same time, key limitations must be acknowledged. The inferred dynamical ages depend sensitively on the accuracy of the intrinsic baryonic mass profiles, including uncertainties in stellar mass-to-light ratios and gas distributions. In addition, the assumption that the evolving baryonic Tully–Fisher relation applies locally as a function of radius remains to be tested in larger samples and controlled simulations.

The present study should therefore be regarded as a proof-of-concept demonstration. Future work will extend the analysis to larger galaxy samples, incorporate improved baryonic mass modelling, and perform systematic comparisons with $\Lambda$CDM and MOND-based frameworks. Cross-validation with independent chronometers, including stellar ages and metallicity-based formation times, will be essential to establish the robustness of the method.

In summary, this work shows that galaxy rotation curves contain previously unexploited temporal information. The radial dynamical chronometer provides a new observational pathway for reconstructing galaxy assembly histories and offers a promising tool for testing theories of gravity and structure formation.

**Acknowledgements**

We thank the SPARC team for making high-quality rotation curves publicly available, and the VENGA, THINGS, and Gemini surveys for their legacy data.

**Data Availability**

The rotation curve data for all galaxies can be obtained from the SPARC database (http://astroweb.cwru.edu/SPARC/). The code used in this analysis is available upon request.